\documentclass[aps,twocolumn,nofootinbib,superscriptaddress,amsfonts,floatfix]{revtex4-1} 
\usepackage{graphicx,amsmath,amssymb,amstext}
\usepackage{amssymb,amsbsy,amsfonts,amsthm,color}
\usepackage{epsfig}
\usepackage{graphicx}
\usepackage{subfigure}
\usepackage{hyperref}
\usepackage{cancel}
\usepackage{balance}
\usepackage{float}

\graphicspath{{Figures/}}

\usepackage{color}
\usepackage[dvipsnames]{xcolor}

\newcommand{\mpl}{M_\mathrm{Pl}}
\newcommand{\mbh}{M_\mathrm{BH}}
\newcommand{\rsh}{r_\mathrm{shell}}
\newcommand{\rDM}{\rho_\mathrm{DM}}
\newcommand{\grchombo}{\mathtt{GRChombo}}
\newcommand{\be}{\begin{equation}}
\newcommand{\ee}{\end{equation}}
\newcommand{\q}{\quad}
\newcommand{\eqn}[1]{Eqn. (\ref{#1})}

\newcommand{\rdm}{\rho_{\mathrm{DM}}}
\newcommand{\dalemb}{\mathop{}\!\mathbin\Box}
\begin{document}
{\hfill KCL-PH-TH/2021-65}
\title{Primordial black hole formation with full numerical relativity}

\author{Eloy de Jong}
\email{eloy.dejong@kcl.ac.uk}

\author{Josu C. Aurrekoetxea}
\email{j.c.aurrekoetxea@gmail.com}

\author{Eugene A. Lim}
\email{eugene.a.lim@gmail.com}

\affiliation{Theoretical Particle Physics and Cosmology Group, Physics Department, Kings College London, Strand, London WC2R 2LS, United Kingdom}

\begin{abstract}
We study the formation of black holes from subhorizon and superhorizon perturbations in a matter dominated universe with 3+1D numerical relativity simulations. We find that there are two primary mechanisms of formation  depending on the initial perturbation's mass and geometry -- via \emph{direct collapse} of the initial overdensity and via \emph{post-collapse accretion} of the ambient dark matter. In particular, for the latter case, the initial perturbation does not have to satisfy the hoop conjecture for a black hole to form. In both cases, the duration of the formation the process is around a Hubble time, and the initial mass of the black hole is $\mbh \sim 10^{-2} H^{-1} \mpl^2$. Post formation, we find that the PBH undergoes rapid mass growth beyond the self-similar limit $M_{BH}\propto H^{-1}$, at least initially. We argue that this implies that most of the final mass of the PBH is accreted from its ambient surroundings post formation.

\end{abstract}
\maketitle

\section*{Introduction} \label{sect:intro}

Primordial black holes (PBHs) form in the early stages of the universe, and their idea was first conceived in the late sixties and early seventies \cite{Carr:1974, Hawking:1971,Zeldovich:1967}. It is notable that it was the potential existence of small black holes from primordial origin that led Hawking to theorize black hole evaporation \cite{Hawking:1974}. It was realised shortly after that PBHs could constitute a significant part of cold dark matter \cite{CHAPLINE1975}, and interest in PBHs has spiked in the recent past as a result. Evaporating PBHs have been suggested as explanations for galactic and extra-galactic $\gamma$-ray backgrounds, short $\gamma$-ray bursts and anti-matter in cosmic rays \cite{Page:1976wx,Carr:1976zz,Wright:1995bi,Lehoucq:2009ge,Kiraly1981,MacGibbon:1991vc,Cline:1996zg} and PBHs could provide seeds for the formation of supermassive black holes and large-scale structure \cite{Carr:1984cr,Bean:2002kx}. Moreover, PBHs could be responsible for certain lensing events \cite{Hawkins:1993,Hawkins:2020}, with recent analysis suggesting that the population of black holes (BHs) detected by the LIGO/Virgo/KAGRA (LVK) observatories \cite{LIGOScientific:2020ibl} may be primordial \cite{LIGOScientific:2020ufj,Franciolini:2021tla}. Additionally, work is underway to use next generation gravitational wave experiments to detect PBH formation and mergers \cite{kozaczuk2021signals,ng2021singleeventbased}. Results obtained by the NANOGrav Collaboration \cite{Arzoumanian_2020} have been associated to PBHs, as well \cite{DeLuca:2020agl,Vaskonen:2020lbd,Kohri:2020qqd,Domenech:2020ers}.

Various formation mechanisms could be relevant for PBHs \cite{Carr:2020gox, Green:2020jor}. These mechanisms include the formation of PBHs during inflation \cite{Clesse_2015,Inomata_2017,Garc_a_Bellido_2017,Ezquiaga:2017fvi}, the collision of bubbles that result from first order phase transitions \cite{Crawford1982,Hawking:1982,Kodama:1982,Leach:2000ea,Moss:1994,Kitajima:2020kig,Khlopov:1998nm,Konoplich:1999qq,Khlopov:1999ys,Khlopov:2000js,Kawana:2021tde,Jung:2021mku}, the collapse of cosmic strings \cite{Kibble:1976sj,Hogan:1984zb,Hawking:1987bn,Polnarev:1991,Garriga:1993gj,Caldwell:1995fu,
MacGibbon:1997pu,Wichoski:1998ev,Hansen:1999su,Nagasawa2005,Carr:2009jm,
Bramberger:2015kua,Helfer:2019,Bertone:2019irm,James-Turner:2019ssu,Aurrekoetxea:2020tuw,Jenkins:2020ctp,Blanco-Pillado:2021klh}, the collapse of domain walls produced during a second order phase transition \cite{Dokuchaev:2004kr,Rubin:2000dq,Rubin:2001yw,Garriga:2015fdk,Deng:2016vzb,Liu:2019lul}, the collapse of a scalar condensate in the early universe \cite{Cotner:2016cvr,Cotner:2017tir,Cotner:2018vug,Cotner:2019ykd} and specific baryogenesis scenarios \cite{Dolgov:1992pu,Dolhov:2020hjk,Green_2016,Dolgov_2009,Kannike_2017}. However, the mechanism that is most relevant for this work is the collapse of overdense regions that are present in the early universe \cite{Carr:1975,Nadezhin:1978,Bicknell:1979, Choptuik:1993,Evans:1994,Niemeyer_1998:nj,Green_1999:gl,Musco:2012au,Yoo:2020lmg}, which may originate from e.g. pre-inflation quantum fluctuations \cite{Carr:1993cl,Carr_1994:cgl,Hodges:1990hb,Ivanov:1994inn,Garcia:1996glw,Randall:1996rsg,Taruya_1999,Bassett_2001}.

In the standard picture, these fluctuations collapse post inflation, while the universe is dominated by radiation energy. The nonzero radiation pressure resists collapse, meaning that the inhomogeneities must be fairly large for PBHs to form. 

It was suggested early on, by using a Jeans length approximation, that an overdensity $\delta$ must be larger than a critical value $\delta_c$ equal to $\frac{p}{\rho} = 1/3$ if PBHs are to form \cite{Carr:1975}, a statement that was checked analytically and numerically soon after \cite{Nadezhin:1978,Bicknell:1979,Novikov:1980}. More recently, analytic and numerical studies have shown that this threshold depends on the initial shape of the overdensity, and can range from $\delta_c = 0.4$ to $0.66$ \cite{Niemeyer_1999:nj,Shibata1999BlackHF,Hawke_2002,Musco:2004ak,Polnarev_2007, Harada:2013epa,Musco:2013, Musco:2019, Musco:2020jjb,Escriva:2020tak}.

PBH formation in matter dominated epochs has also been extensively studied analytically and semi-analytically. In various non-standard universe histories, inflation is followed by a period of matter domination \cite{Khlopov:1985kmz,Carr_2018:cdow,Martin_2020,Allahverdi_2021}. PBH formation in such an early epoch of matter domination was considered early on \cite{Khlopov:1980mg}. More recently, a threshold amplitude for the collapse of a scalar field overdensity was found \cite{Hidalgo:2017dfp}, the effects of non-sphericity \cite{Harada:2016mhb} and inhomogeneity \cite{Kokubu:2018fxy} on the collapse were investigated, the resulting spin of the PBHs was studied \cite{Harada:2017fjm}, the duration of an early epoch of matter domination was constrained by considering the PBH abundance \cite{carrion2021complex} and constraints on the amplitude and spectral index of the collapsing scalar field were obtained \cite{Carr:2017edp}.

\begin{figure*}[t]
    \includegraphics[width=\linewidth]{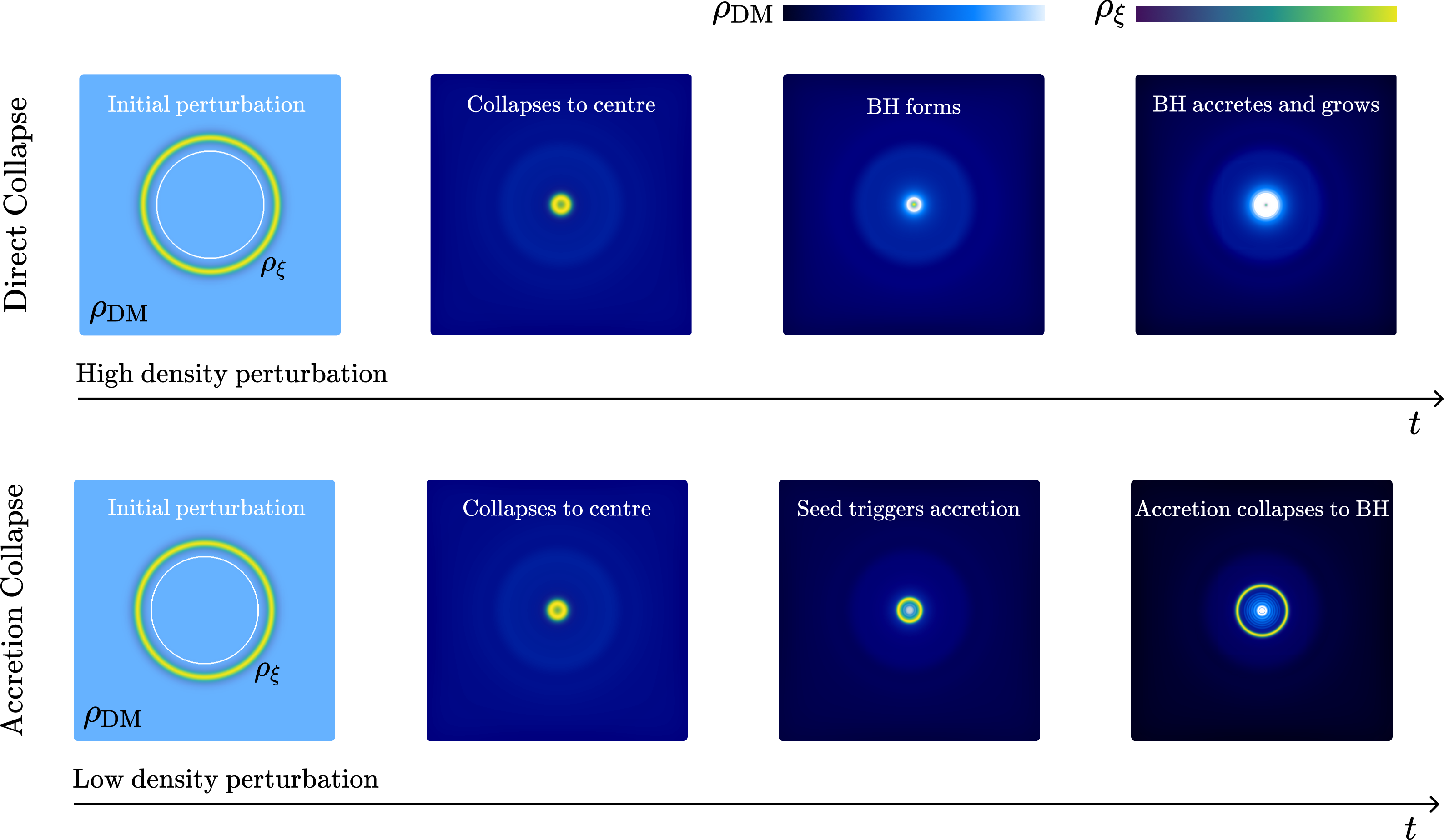}
    \caption{\textbf{Direct collapse and accretion driven mechanisms:} The figure summarizes the two distinct processes of PBH formation studied in this paper. 
    Top panel shows the \textit{direct collapse} mechanism, where a high density initial superhorizon perturbation collapses and forms a black hole as soon as the perturbation reaches the centre.
    Bottom panel depicts the \textit{accretion driven collapse} mechanism, where a lower density initial perturbation collapses and acts as a seed to trigger the accretion of the background dark matter, which subsequently collapses to form a black hole. Both start from the same initial radius $R_0$, but with different initial amplitudes $\Delta \xi$.   In the leftmost figures, we show the initial size of the Hubble horizon (white solid line), which will grow as time evolves. In the other figures, the Hubble horizon has grown larger than the box size. Colourbars are shown in the top right, with lighter (darker) colours signifying higher (lower) energy densities, and scales fixed per mechanism. Video comparisons of these mechanisms can be found \href{https://youtu.be/fqEBlCybF8I}{here} \cite{Movie1} and \href{https://youtu.be/4N5e2RnUkmU}{here} \cite{Movie2}.} 
    \label{fig:panel_summary}
\end{figure*}

In this work, we use full 3+1D numerical relativity simulations to investigate the collapse of subhorizon and superhorizon non-linear perturbations in an expanding universe that is dominated by matter. We model the expanding background and the collapsing perturbation using an oscillating massive scalar field and massless scalar field respectively. The massless scalar field's initial energy is thusly contained purely in its gradients. We will show that there are two broad mechanisms of black hole formation -- via \emph{direct collapse} for the case where the overdensity is sufficiently large that it will form a black hole, and via \emph{post-collapse accretion} for the case where the overdensity is smaller. In both cases, the process is rapid and its duration is around a single Hubble time, forming PBHs with initial masses of $\mbh H\sim 10^{-2}\mpl^2$. We illustrate these mechanisms in Fig. \ref{fig:panel_summary}.

Our choice of fundamental scalar fields as dark matter, instead of a pressureless cold fluid (as suggested by \cite{East:2019chx}), is prompted by our focus on an early time (i.e. pre-BBN) matter dominated phase instead of the present late time matter dominated phase. Such early era matter domination is often driven by a non-thermal fundamental scalar or moduli dynamics \cite{Acharya:2008bk}  instead of the more familiar cold pressureless fluid such as thermal WIMP dark matter. Furthermore, early matter phases will eventually transition into a radiation domination epoch, such that the standard Hot Big Bang cosmological evolution can proceed. Such a phase transition from matter domination into radiation domination can then be achieved through the decay of the scalar field into either standard model particles or intermediaries.

This paper is organised as follows. In section \ref{sect:setup} we explain the numerical setup we use to evolve a scalar perturbation in a dark matter dominated background. In section \ref{sect:bhformation} we introduce the two aforementioned formation mechanisms and their characteristics, and we study the properties of the black holes that are formed post collapse. In section \ref{sect:mass_distribution} we comment on the post formation growth of the PBHs, and we conclude in section \ref{sect:conclusions}.

\section{Early Matter Domination Epoch with Scalar fields} \label{sect:setup}

We will use a metric with $-+++$ signature, in Planck units $\hbar=c=1$ such that $G=\mpl^{-2}$.  The action we will consider is 
\begin{equation}
S = \int ~d^4 x \sqrt{-g}\Big[\frac{\mpl^2}{16\pi}R - \mathcal{L}_\phi - \mathcal{L}_\xi\Big]~,
\end{equation}
involving a massive scalar field $\phi$ with mass $m$ that models the ambient dark matter, and a massless scalar field $\xi$ that sources the initial perturbation. They are both minimally coupled to gravity but not otherwise coupled to one another, i.e.
\begin{align}
\mathcal{L}_\phi &=\frac{1}{2}\nabla_{\mu}\phi\nabla^\mu\phi + \frac{m^2 \phi^2}{2}~,~\mathrm{and}~ \\
\mathcal{L}_\xi &= \frac{1}{2}\nabla_{\mu}\xi\nabla^\mu\xi ~.
\end{align}
Since the field $\xi$ has no potential, it will only influence dynamics via its gradients. Furthermore, it will dilute much more rapidly than dark matter, and hence it does not affect the long term dynamics of the system once its initial job of sourcing a perturbation is done\footnote{In principle, we could use a single massive scalar $\phi$. However, in practice, we find that  large perturbations of the massive scalar would introduce a large infusion of potential energy into the dynamics of the background resulting in non-matter dominated evolution, at least initially.}.

When the gradients in $\xi$ are negligible, the spacetime dynamics are dominated by the behaviour of the background scalar field $\phi$. When $\phi$ is additionally homogeneous on a given spatial hyperslice, the metric of the spacetime is well described by the Friedman-Lema\^{i}tre-Robertson-Walker (FLRW) line element
\begin{equation}
ds^2 = -dt^2 + a(t)^2(dr^2+r^2d\Omega^2_2)~,
\end{equation}
where $d\Omega^2_2 =d\theta^2 + \sin^2\theta d\phi^2$. The scale factor $a(t)$ evolves according to the Friedmann equation $H^2 = 8\pi\rho/3\mpl^2$, where $H(t)\equiv \dot{a}/a$ is the Hubble parameter\footnote{Dotted variables are derivatives with respect to cosmic time $t$.}. The equation of motion for $\phi$ reduces to the Klein-Gordon equation
\begin{equation} \label{KGphi}
  \ddot{\phi} + 3H\dot{\phi} + \frac{dV}{d\phi} = 0~,
\end{equation}
where the Hubble parameter is
\begin{equation}\label{eq:friedmann}
H^2 \equiv \frac{8\pi}{3\mpl^2}\left(\frac{1}{2}\dot{\phi}^2 + V(\phi)\right)~,
\end{equation}
and the corresponding pressure is given by
\begin{equation}\label{eos}
  p_{\mathrm{DM}} = \frac{1}{2}\dot{\phi}^2 - V(\phi)\,.
\end{equation}
If the oscillation of $\phi$ is sufficiently undamped, which is the case if $2m \gg 3H$ \footnote{This condition is obtained by interpreting \eqn{KGphi} as the equation of motion for a damped oscillator of the form $m\frac{dx^2}{dt^2} + c\frac{dx}{dt} + kx = 0$. The condition for undamped oscillation then becomes $c^2 - 4mk < 0$ or $9H^2 - 4m^2 < 0$. }, the friction term in \eqn{KGphi} can be neglected. The dynamics of $\phi$ are then approximately given by a simple harmonic oscillator $\phi(t) = \phi_0\cos{\big(mt\big)}$, whose pressure is

\begin{equation}
    p_{\mathrm{DM}} = \frac{\phi_0^2 m^2}{2}\Big(\sin^2{\big(mt\big)} - \cos^2{\big(mt\big)}\Big)~.
\end{equation}
As long as the oscillation period $T$ is sufficiently smaller than one Hubble time, this averages to zero over one Hubble time, i.e. $\langle p_{\mathrm{DM}} \rangle =0$, resulting in a dark matter dominated expansion, which can be interpreted as a model for pressureless dust \cite{gu2007oscillating} at large scales.

Meanwhile, the massless scalar field $\xi$ provides the energy density perturbation that will trigger BH formation. In this paper, we exclusively consider initially static spherically symmetric perturbations and we leave the generalisation to fewer degrees of symmetry for future work. We choose the initial configuration of $\xi$ to be space dependent as
\begin{equation} \label{xi_profile}
    \xi(t=0,r) = \Delta\xi ~ \tanh{\Big[\frac{r - R_0}{\sigma}\Big]},
\end{equation}
where $\Delta\xi$, $\sigma$ and $R_0$ are the amplitude, width and the initial size of the perturbation respectively. We comment further on this perturbation shape in appendix \ref{section:initialdata}. The mass of the initial perturbation scales roughly as $R_0^2$.  We emphasise that this perturbation is non-linear, despite its moniker. Nevertheless, its massless nature means that it will propagate very close to the speed of light. 

The initial staticity of the perturbation reflects the limited dynamics of a superhorizon perturbation that is frozen out. However, frozen out does not imply motionless. Nevertheless, the $\xi$ field is massless and accelerates rapidly to velocities around $c$, regardless of its initial velocity. Thus, combined with the fact that a static configuration simplifies the constraints the initial data must satisfy (which is discussed more detail in appendix \ref{section:initialdata}), we deem this an acceptable simplification.

Given the initial static configuration, we expect to see the perturbation split into an infalling mode, which drives the PBH formation, and an outgoing mode, which rapidly disperses. Our simulation box has periodic boundary conditions -- we ensure that the simulation domain is sufficiently large that the outgoing mode does not reach the boundary before PBH formation takes place. The background scalar field $\phi$ starts from rest, so that $\dot{\phi}=0$ and the initial Hubble parameter in the absence of inhomogeneities is $H_0^2 = 8\pi \mpl^{-2} V(\phi_0)/3$ via \eqn{eq:friedmann}.

Since the configuration of $\xi$  breaks the homogeneity of the initial spatial hyperslice, to set up the correct initial conditions for the metric, we will solve the Hamiltonian constraint. We choose a conformally flat ansatz for the 3-metric $\gamma_{ij}$,
\begin{equation}
dl^2 = \psi^{4}(dx^2+dy^2+dz^2)\,.
\end{equation}
Then, the Hamiltonian constraint reduces to an equation for the conformal factor $\psi$
\begin{equation} \label{eq:hamconstraint}
    \mathcal{H} = \dalemb\psi - \frac{\psi^5}{12}K^2 + 2\pi\mpl^{-2}\psi^5\rho = 0\,,
\end{equation} 
where 
\begin{equation}
\rho = \rho_\xi + \rho_{\mathrm{DM}} = \frac{\psi^{-4}}{2}\left(\partial_i\xi\right)^2 + V(\phi_0) ~.
\end{equation}
Here the local expansion $K$ is the trace of the extrinsic curvature, $K = \mathrm{Tr} K_{ij}$.  \eqn{eq:hamconstraint} then becomes 
\begin{equation} 
\dalemb\psi - \frac{\psi^5}{12}\left(K^2 - 9H_0^2\right) + \pi\mpl^{-2}\psi\left(\partial_i\xi\right)^2 = 0\,.
\end{equation}
We choose an initially expanding spacetime with $K = -3H_0$, so that the periodic integrability condition is satisfied\footnote{The initial energy density of the system is completely dominated by the scalar potential of the homogeneous dark matter field $\phi$, which allows us to neglect the perturbation field $\xi$. Then, the main contribution to the initial energy density is given by the (homogeneous) value of the potential and thus $K^2=24\pi V(\phi_0)$.} \cite{Bentivegna:2013xna,Yoo:2018pda,Clough:2016ymm}. The eventual Hamiltonian constraint only depends on the radial coordinate $r$ due to the spherical symmetry of the setup, and we solve for the conformal factor $\psi$ numerically
\begin{equation} \label{eq:ham_constraint}
\frac{\partial^2\psi}{\partial r^2} + \frac{2}{r}\frac{\partial\psi}{\partial r} - \frac{\psi^5}{12}\left(K^2 - 9H_0^2\right) + \frac{\pi\psi}{\mpl^{2}}\left(\frac{\partial\xi}{\partial r}\right)^2 = 0\,.
\end{equation}

\section{Primordial black hole formation} \label{sect:bhformation}

Our main scale of reference will be the initial size of the unperturbed  Hubble horizon $H_0$, which is fixed for all simulations by choosing the initial value of the scalar field $\phi$ to be $\phi_0=7.8\times 10^{-3}\mpl$, with $m\approx 10^2 H_0$. In the following, we will vary the initial size of the perturbation from subhorizon to superhorizon, $R_0 H_0 \in \left[0.575,~1.6\right]$. We will also vary the perturbation amplitude within the range $\Delta\xi\mpl^{-1} \in \left[0.075,~0.12\right]$, whilst keeping the initial width fixed to $\sigma_0=0.15 H_0^{-1}$, such that the ratio between the maximum gradient energy density to dark matter energy density is $\rho_{\xi}/\rho_\mathrm{DM}\sim 1$. As the perturbation mass is small compared to the Hubble mass in all scenarios we consider, the background expansion is not significantly influenced by the perturbation's presence.

Our simulations reach just into the superhorizon regime and at the moment, we are limited from exploring this regime further numerically by the computational costs of such simulations. Nonetheless, we will argue in section \ref{sect:accretion} that our simulations already capture some superhorizon dynamics. Moreover, we argue that it may not be necessary to probe the superhorizon regime much further, as our model is effectively pressureless. The corresponding speed of sound is therefore small, and physics already propagate very slowly on scales just past the horizon size.

We find that, for both subhorizon and superhorizon perturbations, PBH formation occurs via two possible mechanisms -- a \emph{direct collapse} mechanism whereby the PBH is formed by the initial perturbation of $\xi$ itself, and a \emph{post-collapse accretion} mechanism whereby the initial perturbation of $\xi$ sources a gravitational potential that then accretes the background dark matter  $\phi$ until a PBH forms. What determines the type of PBH formation process depends (unsurprisingly) on both the geometry and mass of the initial perturbation shell, as well as the expansion rate of the background cosmology. We will discuss these two mechanisms below \footnote{In this paper, we have used the background energy density, or equivalently the background scale factor, as time. This corresponds to the cosmic time infinitely far away from the centre of the PBH. However, in numerical relativity simulations, the foliation of spatial hyperslices is dynamically driven by the so-called puncture gauge, which is required to enforce numerical stability in the presence of future singularities. In that context, we have assumed that the mass of a PBH is identified with its foliation, when in principle one should identify it via null geodesics from the black hole horizon to infinity.  In other words, what the cosmological observer (with their own clocks tuned to cosmic time) far away from the black hole deduces as the properties of the black hole, e.g. its mass, should be information that is emitted (by light or GW) from the black hole and then propagated to this observer. Our approximation assumes this ``time lag'' between the local time (i.e. foliation time) and the cosmic time is negligble. This inaccuracy should be minor and would not affect the main conclusions of the paper. }.

\subsection{Direct collapse} \label{sect:direct_collapse}

\begin{figure}[t]
    \includegraphics[width=\linewidth]{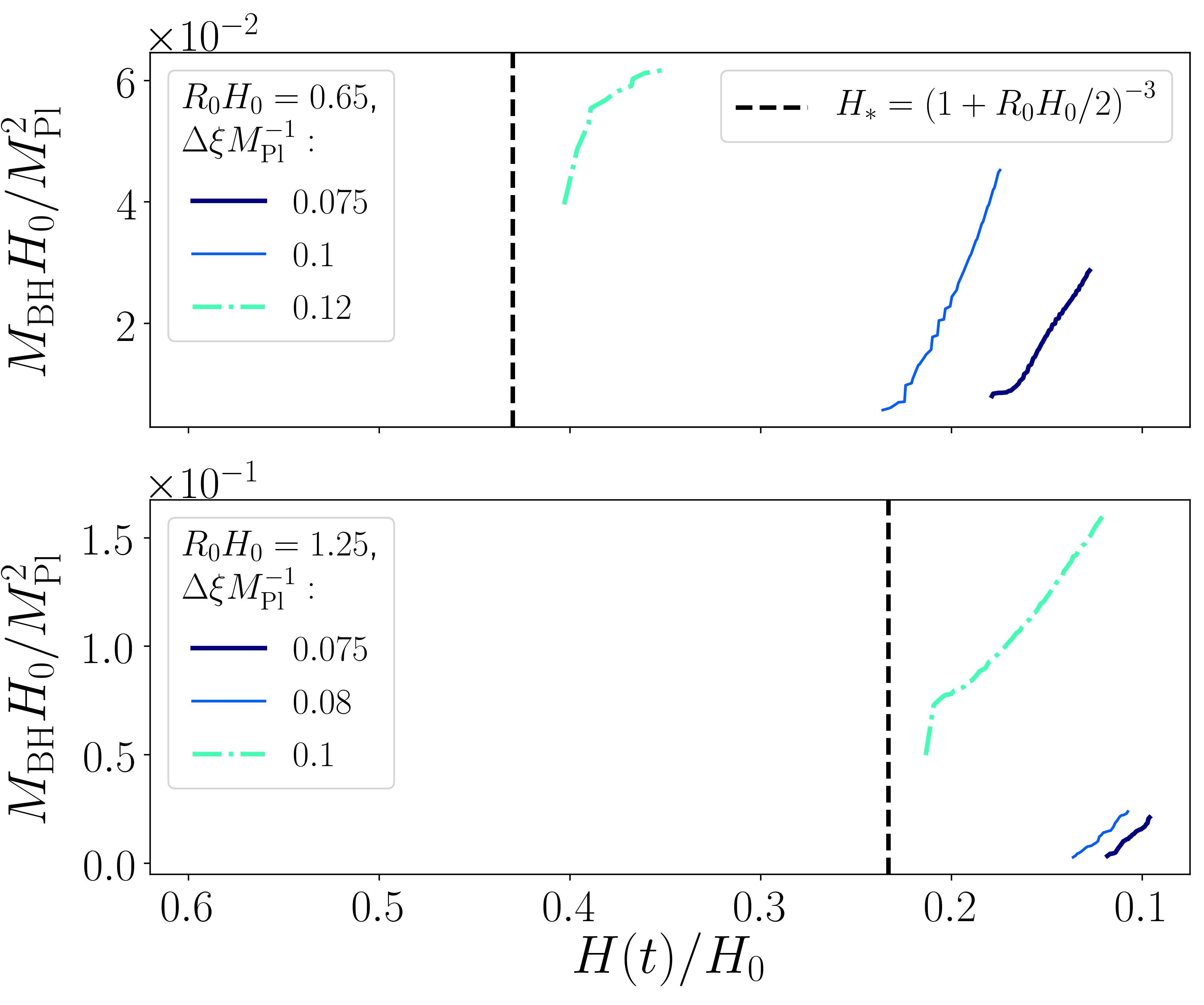}
    \caption{\textbf{Black hole formation for different perturbation amplitudes.} The top (bottom) panel shows mass of the formed BHs as a function of the Hubble parameter $H(t)$ at infinity, for subhorizon (superhorizon) collapse respectively. Vertical dashed black lines correspond to the time at which the perturbation reaches the centre according to \eqn{eqn:astar}. BHs formed through direct (accretion) collapse are shown in dash-dotted (solid) lines. For accretion collapse BHs, increasing the amplitude $\Delta\xi$ makes that the BH forms earlier with a smaller initial mass. Our simulations are in good agreement with the hoop conjecture prediction that the threshold is $\Delta\xi\mpl^{-1}\approx 0.1$ for $R_0H_0 =0.65$ and $\Delta\xi\mpl^{-1}\approx 0.07$ for $R_0H_0 =1.25$. In direct collapse, part of the collapsing perturbation ends up within the black hole, corresponding to a larger initial mass.} 
    \label{fig:figure_A1}
\end{figure}

In the direct collapse scenario, the perturbation collapses towards its geometric centre (to which we will henceforth simply refer as the centre) and forms a black hole directly on its own, without significant accretion of the background DM density. We will now estimate the time a shell takes to undergo direct collapse. The perturbation field $\xi$ is massless, and hence if we ignore the backreaction of the shell on the background geometry, it propagates along null-like geodesics \footnote{If the perturbation field $\xi$ instead has mass $m_{\xi} \sim m_{\phi}$, then the collapse time is roughly the free fall time $\tau_{\mathrm{ff}} \sim \sqrt{\mpl/m_{\xi}\xi}$ which is also roughly one Hubble time.}. In an FLRW background, the scale factor $a$ is given by the null element $dt^2 =a^2(t)dr^2$. Solving this kinematic equation, the co-moving radius of the shell is then
\begin{equation}
\rsh = R_0a_0^{-1}-2H_0^{-1}a_0^{-1}\left[\left(\frac{a}{a_0}\right)^{1/2}-1\right]~,\label{eqn:shell_trajectory}
\end{equation}
where we set $a_0\equiv 1$ to be the initial scale factor at the initial time. The value of the scale factor at the moment the shell collapses to the centre $a_*$ is then the solution to the equation $\rsh(a_*)=0$, namely
\begin{equation}
a_* = a_0\left[1+ \frac{R_0H_0}{2}\right]^2~,\label{eqn:astar}
\end{equation}
which is roughly a Hubble time. 
Notice that $a_*$ is independent of the initial mass and depends only on $R_0$. We show in Fig. \ref{fig:figure_A1} that this analytic estimate is in good agreement with our numerical results.

To determine whether or not a given initial perturbation shell will undergo direct collapse into a black hole, we consider the \emph{width} of the shell at the time when the shell reaches the centre $\sigma_* = \sigma(a_*)$. Ignoring backreaction again, since the field $\xi$ is massless, the width of the shell as it collapses towards the centre scales as the expansion rate, i.e. 
\begin{equation}
\sigma(a) = \sigma_0 a ~. \label{eqn:width_eom}
\end{equation}
Thus the width of the shell when it reaches the centre is simply $\sigma(a_*) = \sigma_0a_*$. At this moment, applying the hoop conjecture\footnote{The presence of an expanding background modifies the hoop conjecture somewhat in general \cite{Saini:2017tsz}, but we checked that the effects are negligible in our analytic estimates. } suggests that if the condition
\begin{equation}
\sigma(a_*) < 2GM_{\mathrm{infall}}~,\label{eqn:hoop}
\end{equation}
is satisfied,  where $M_{\mathrm{infall}}$ is half\footnote{The infalling mass is half the initial mass, since the other half will radiate outwards to infinity, so the shell's initial (vanishing) momentum is conserved.}
the initial mass of the shell obtained by integrating the gradient energy of the profile $\xi(r)$ roughly given by  \eqn{xi_profile} in flat space
\begin{equation}
M_{\mathrm{infall}} \approx \frac{1}{2}\int dr~4\pi r^2 \frac{1}{2}\left(\frac{\partial \xi}{\partial r}\right)^2~,\label{eqn:infall_mass}
\end{equation}
then a black hole will form. This result is again in good agreement with our numerical results, as shown in Fig. \ref{fig:figure_A1}.

\begin{figure}[t]
    \includegraphics[width=\linewidth]{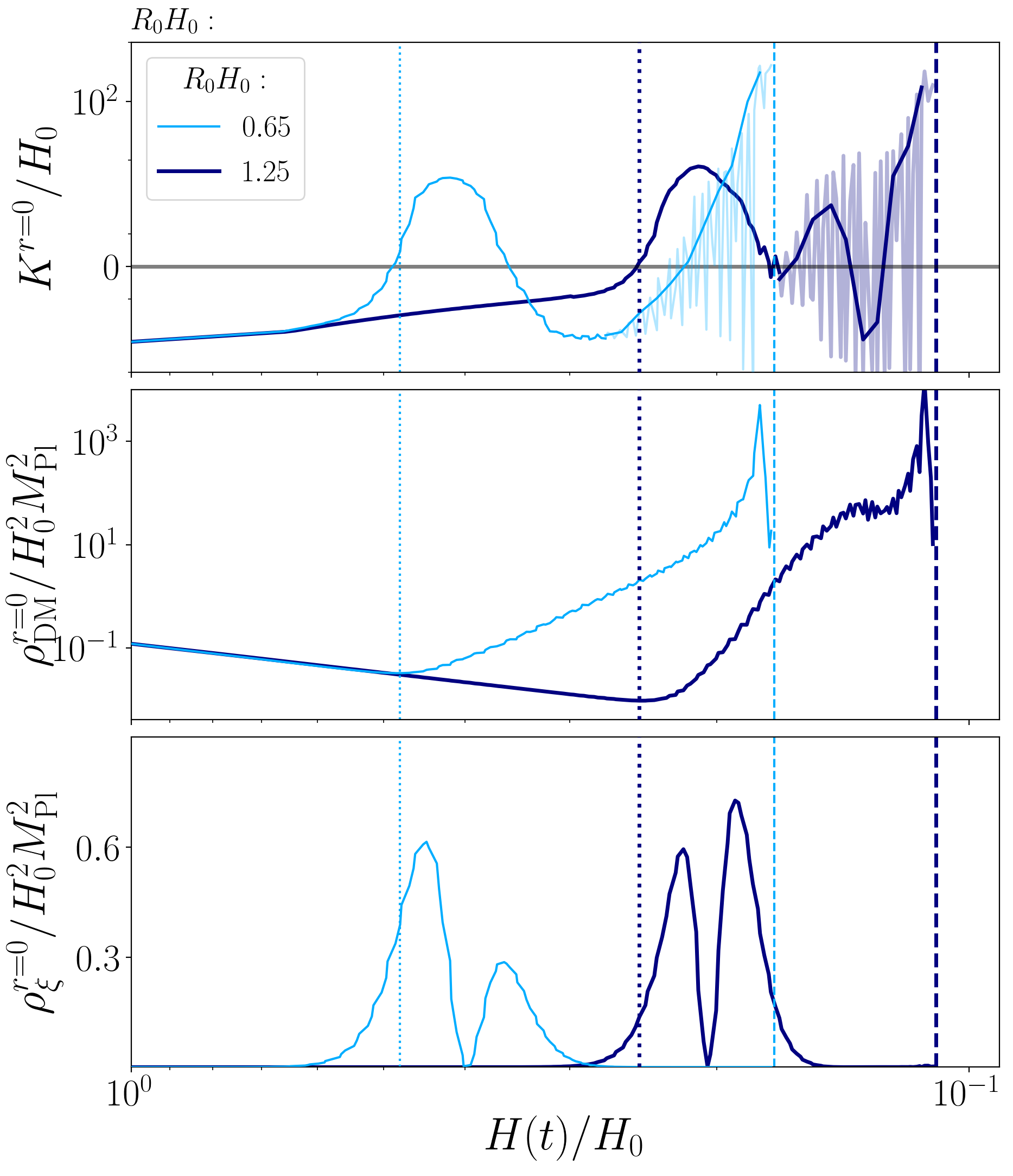}
    \caption{\textbf{Evolution of the local expansion $K$ and energy densities $\rho_{DM}$ and $\rho_\xi$} at the centre of the collapse $r=0$, as a function of the Hubble parameter $H(t)$ at infinity -- recall that $K>0$ corresponds to locally collapsing spacetime.  A representative subhorizon (superhorizon) is shown in thin light blue (thick dark blue) in the {\bf accretion collapse case}. The top, middle and bottom panels show the evolution of the expansion, the background energy density and gradient energy density respectively. Initially, the background energy density decays as $\rho_\mathrm{DM} \sim a(t)^{-3}$. When the perturbation reaches the centre (dotted vertical lines) and disperses, gravitational effects decouple the system and stop the local expansion, acting as a seed for the accretion of the background matter $\rho_\mathrm{DM}$. The accretion of the background matter continues until and after a black hole forms (dashed vertical lines).} 
    \label{fig:rhos_vs_r}
\end{figure}

The fact that such simple estimates agree with our numerical results suggests that the backreaction of the perturbation on the background dynamics is not very important, at least at the level of determining when and how a black hole will form, even if the shell density is large and locally $\rho_\xi > \rDM$. This is backed up by our numerical simulations, where we see that the $\rDM$ profile is not strongly affected by the presence of $\rho_\xi$, at least initially, as can be seen in a video of the numerical evolution of the energy densities \href{https://youtu.be/4N5e2RnUkmU}{here} \cite{Movie2}.

\subsection{Accretion collapse} \label{sect:accretion}

\begin{figure}[t]
    \includegraphics[width=\linewidth]{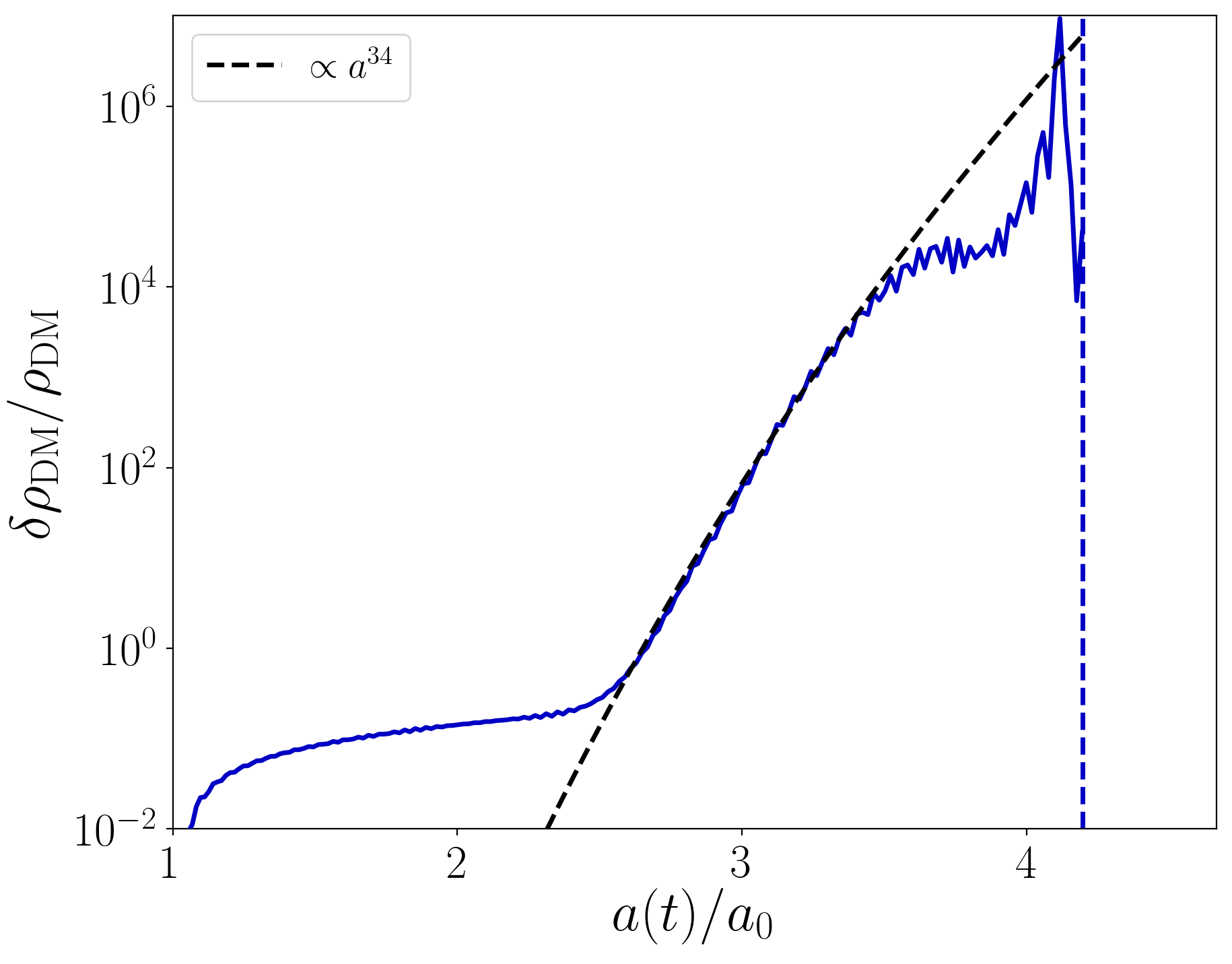}
        \caption{\textbf{Rate of growth for the local dark matter overdensity} $\delta \rho_{\mathrm{DM}}/\rho_{\mathrm{DM}}$ at the centre of the collapse is well beyond the linear approximation as $\delta\rho_\mathrm{DM}/\rho_\mathrm{DM}\propto a^{34}$. Near black hole formation (vertical dashed blue line), the accretion rate tapers off, although some of this tapering effect is due to our gauge condition.}  
    \label{fig:deltarho_pert}
\end{figure}

On the other hand, if $\sigma(a_*) > 2GM_{\mathrm{infall}}$, a black hole does not form directly. In this case, the energy density of the perturbation $\rho_{\xi}$ disperses after reaching the centre and becomes locally sub-dominant to the background energy density $\rdm$. Nevertheless, the presence of $\xi$ generates a gravitational potential well in the centre, which seeds accretion of the background DM and eventually causes a collapse into a black hole. We illustrate this process in Fig. \ref{fig:rhos_vs_r}.

In this phase, the initially homogeneous and expanding background spacetime is made to locally collapse by the perturbation, with the expansion $K$ locally changing sign from negative (expanding) to positive (contracting), decoupling the region near the centre from the rest of the expanding background. The local dark matter begins to accrete at an extremely high rate $\delta \rdm/\rdm \propto a^{34}$, as shown in Fig. \ref{fig:deltarho_pert}. Once sufficient DM mass has accumulated, a PBH forms. This process takes around an e-fold to complete.  This rapid accretion rate is much higher than that predicted from linear theory, which is $\delta \rdm/\rdm  \propto a$, indicating that the process is highly non-linear.

\begin{figure}[t]
    \includegraphics[width=\linewidth]{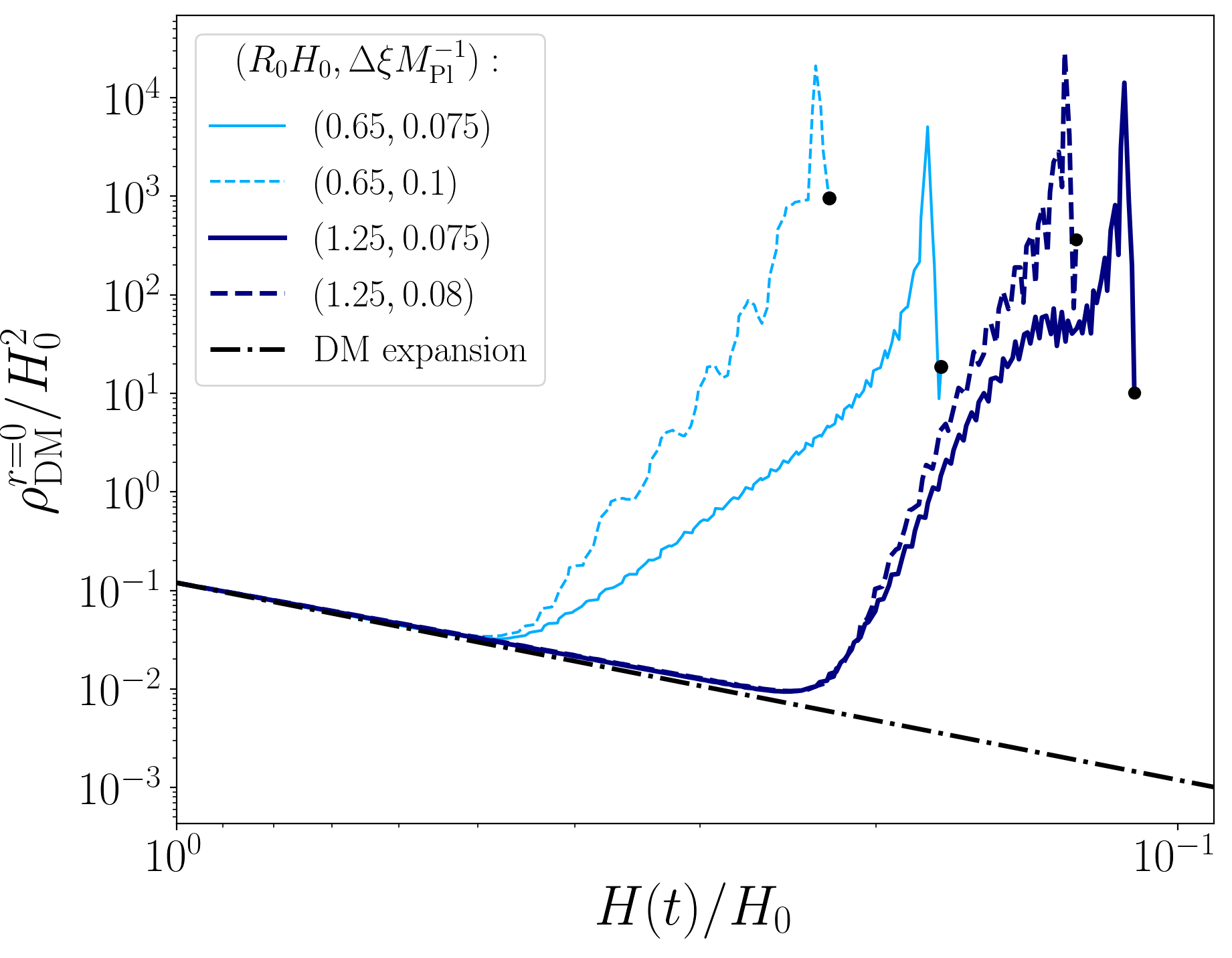}
        \caption{\textbf{Evolution of the dark matter energy density} at the centre of the collapse for a set of initial radii $R_0$ and amplitudes $\Delta\xi$. For same radii perturbations, accretion begins at the same time. However, the accretion rate is larger for larger amplitudes, which results in the formation of a black hole at an earlier time.} 
    \label{fig:rhoDM_amp}
\end{figure}

From our simulations, we note two salient points. Firstly, if we consider shells that undergo accretion collapse, for fixed initial amplitude $\Delta\xi$, the smaller the initial $R_0$ (and therefore the smaller the mass) of the initial perturbation, the more massive the initial mass of the PBH. This somewhat counter-intuitive result is due to the fact that the PBH forms via accreting DM, thus a more massive seed will generate a \emph{steeper} potential well, and hence the Schwarzschild radius is reached earlier and at a smaller value for the PBH mass. To confirm this, we checked that keeping $R_0$ fixed but increasing $\Delta \xi$ also yields a less massive initial PBH -- this is true for both subhorizon and superhorizon cases, as can be seen in Fig. \ref{fig:figure_A1}. In Fig. \ref{fig:rhoDM_amp} we plot the dark matter energy density for two different values for $R_0$ and $\Delta\xi$. We confirm that larger amplitudes (and thus more massive seeds) result in a faster accretion rate. 

Secondly, as $R_0$ approaches $H_0^{-1}$, the expansion rate of the universe begin to exert a competing effect. For shells with larger $R_0$, it takes \emph{longer} for the shell to reach the centre, and thus a smaller $\rho_{\xi}$ and less steep potential when accretion begins. This leads to an increase in the initial mass of the PBH following our argument above -- resulting in the ``bump'' in the initial mass of the black holes (e.g. the black dots in Fig. \ref{fig:mass_vs_rho}).

  \section{PBH Growth and Final Mass} \label{sect:mass_distribution}
\begin{figure}[t!]
    \includegraphics[width=\linewidth]{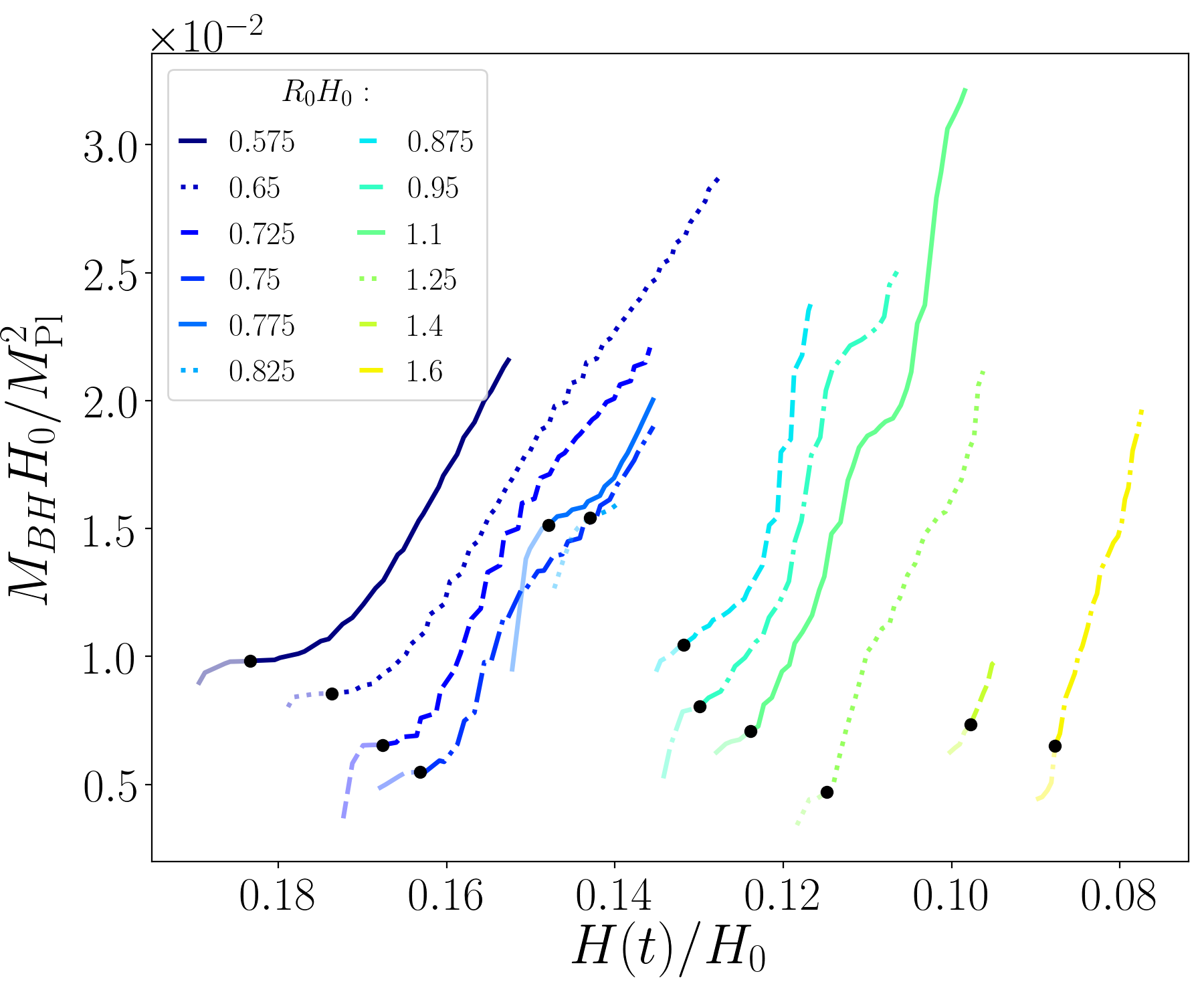}
    \caption{\textbf{Summary of simulations} showing the black hole $M_\mathrm{BH}$ as a function of the Hubble parameter $H(t)$ at infinity, for various initial radii $R_0$, for $\Delta\xi\mpl^{-1} = 0.075$. The growth rate of the black hole mass is larger for larger shells, because they source a larger gravitational potential. Black dots correspond to the initial black hole masses at formation, identified using an apparent horizon finder.}.
    \label{fig:mass_vs_rho}
\end{figure}

In the cases of both direct and accretion collapse, the initial mass of the PBH formed is small compared to the Hubble horizon, $\mbh H_0\sim 10^{-2}\mpl^2$ -- see Fig. \ref{fig:mass_vs_rho}.  Once the initial PBH has formed, the PBH accretes DM from its surroundings in the growth phase at a rate that depends on the steepness of the potential and the density of the surrounding DM ``scalar cloud'' \cite{Hui:2019aqm,Clough:2019jpm,Bamber:2020bpu,Hui:2021tkt}. In general and regardless of the details of the parameters, we find that the initial accretion rate is much higher than the linear theory prediction of $\delta \rdm/\rdm \propto a$, as mentioned above.  This growth rate is roughly constant, at least initially, and its contribution to the mass of the PBH will rapidly dwarf that of its initial mass.  

In a matter dominated universe,  naive Newtonian collapse suggests that the maximum mass of the black hole is bounded by $\mbh H\sim \alpha \mpl^2$ \cite{Carr:1974}, where $\alpha\lesssim 1$ is some constant which depends on the exact details of the accretion. This suggests a self-similar growth at some equilibrium point. In references \cite{Harada:2004pf,Harada:2004pe}, it was demonstrated numerically that while the initial growth can be rapid, it will not achieve self-similar growth as accretion is not efficient once the black hole decouples from the background spacetime. However, these works used a stiff massless scalar field as ambient matter instead of a massive scalar field, which more accurately models the ambient DM. From  Fig. \ref{fig:mass_vs_rho}, we find that $M\sim H^{-\beta} $ where $\beta\gg 1$. As $M$ approaches the Hubble horizon, we expect $\beta \leq 1$ although unfortunately, we were unable to track the growth of PBH beyond a few factors of their initial mass, as the numerical cost becomes prohibitive. 

As long as the universe is dominated by DM, the black hole will continually accrete and grow without end. This would be the case if the PBH is formed in the present late time DM dominated epoch -- however such late time PBH has already been ruled out \cite{Carr:2020gox,Green:2020jor}. As we mentioned in the introduction, we consider instead an early phase of DM domination before transitioning into the era of radiation domination prior to the onset of Big Bang Nucleosynthesis  (BBN), i.e. before the temperature of the universe is around $1~\mathrm{MeV}$. This provides a natural cut-off for the growth of the PBH. 

Nevertheless, if we assume that the rapid growth we observe continues until $\mbh H \sim \mpl^2$, and that the BH grows self-similarly after, it is implied that the final mass of the PBH is independent of when it forms.  This means the final mass of the PBH is given by
\begin{equation}
M_\mathrm{BH} \approx 10^{38} \left(\frac{1~\mathrm{MeV}}{T}\right)^2 g \approx 10^5 \left(\frac{1~\mathrm{MeV}}{T}\right)^2 M_\odot~, 
\end{equation}
where $T$ is the temperature of the universe at the onset of radiation domination. Taking $T_\mathrm{BBN}=1~\mathrm{MeV}$ as the natural cut-off for the growth of the PBH, the most massive black holes that can be formed via this accretion mechanism are $M_\mathrm{BH} \approx 10^{38} g \approx 10^5 M_\odot~$ \cite{Carr:2004kgc,Green:2014faa}.

On the other hand, if the PBH growth asymptotes to a slower rate than  the self-similar rate, or achieve self-similarity before $\mbh \sim H^{-1} \mpl^2$, then our simulations suggests that $\mbh \gtrsim 10^{-2} H^{-1} \mpl^2$, where $H$ is the Hubble parameter when the PBH forms. This means that PBH formed around $T\sim 5$ MeV, $ \mbh \gtrsim  40M_\odot$ could form the basis of the  population of massive BH that are being detected today by the LVK observatories.

  \section{Summary and Discussion} \label{sect:conclusions}

In this paper, we demonstrate that superhorizon non-linear perturbations can collapse and form PBHs in a matter dominated universe, using full numerical relativity. We show that, depending on the mass of the initial perturbation shell, this happens via either the direct collapse or the accretion collapse mechanisms. We provide an analytic criterion  \eqn{eqn:hoop} using the hoop conjecture to determine which mechanism is relevant for a given setting, and compute the timescale of collapse using the same prescription. Despite the ${\cal O}(1)$ non-linearity, we find that the dynamics of collapse can be modeled as a simple superhorizon mass shell collapsing in an expanding background. This suggests that semi-analytic estimates of PBH formation in a matter dominated era are broadly accurate. 

On the other hand, details matter. We showed that  even in the cases where the perturbation is insufficient on its own to form a PBH in a direct collapse, non-linear accretion rates are far higher than what standard linear theory predicts, causing a rapid collapse into a PBH via accretion of ambient DM.  In both the direct collapse and accretion collapse formation cases, the initial mass of the PBH is roughly $\mbh \sim 10^{-2}H^{-1}\mpl^2$, but formation is followed by an extremely rapid growth $M\propto H^{-\beta}$ where $\beta\gg 1$. Presumably, this growth will asymptote to either the self-similar rate $\beta=1$ or the decoupled rate $\beta <1$ \cite{Harada:2004pf,Harada:2004pe}.

Interestingly, even if the self-similar rate is not achieved, the fact that most of the mass of the PBH is gained through post-formation accretion suggests that there might be a mechanism to generate PBHs with non-trivial spin.  Such non-trivial spin might for example be generated by the collapse of a non-spherically symmetric shell, even if the shell is initially spinless. In that case, the PBH might not form in the centre of the initial mass distribution and thus form with spin, whilst outgoing radiation carries away angular momentum of opposite sign, such that angular momentum is still globally conserved, as suggested by \cite{Harada:2016mhb, Harada:2017fjm}. We will explore this possibilty in a future publication.\\

\acknowledgments
We acknowledge useful conversations with Bernard Carr, Katy Clough,  Pedro Ferreira, Tiago Fran\c{c}a, Cristian Joana, David Marsh, John Miller, Yolanda Murillo, Miren Radia and Chul-Moon Yoo. We would also like to thank the GRChombo Collaboration team \href{http://www.grchombo.org}{(http://www.grchombo.org/)} and the COSMOS team at DAMTP, Cambridge University for their ongoing technical support.  This  work  was  performed  using  the Leibnitz Supercomputing Centre SuperMUC-NG under  PRACE grant Tier-0 Proposal 2018194669, on the J\"ulich Supercomputing Center JUWELS HPC under PRACE grant Tier-0 Proposal 2020225359, COSMA7 in Durham and Leicester DiAL HPC under DiRAC RAC13 Grant ACTP238, and the Cambridge Data Driven CSD3 facility which is operated by the University of Cambridge Research Computing on behalf of the STFC DiRAC HPC Facility.  The  DiRAC  component  of CSD3 was funded by BEIS capital funding via STFC capital grants ST/P002307/1 and ST/R002452/1 and STFC operations grant ST/R00689X/1.  

\bibliography{mybib}

\clearpage
\appendix

\section{Numerical methodology} \label{appendix:initialdata}

\subsection{Evolution equations}

\begin{figure}[t]
    \includegraphics[width=\linewidth]{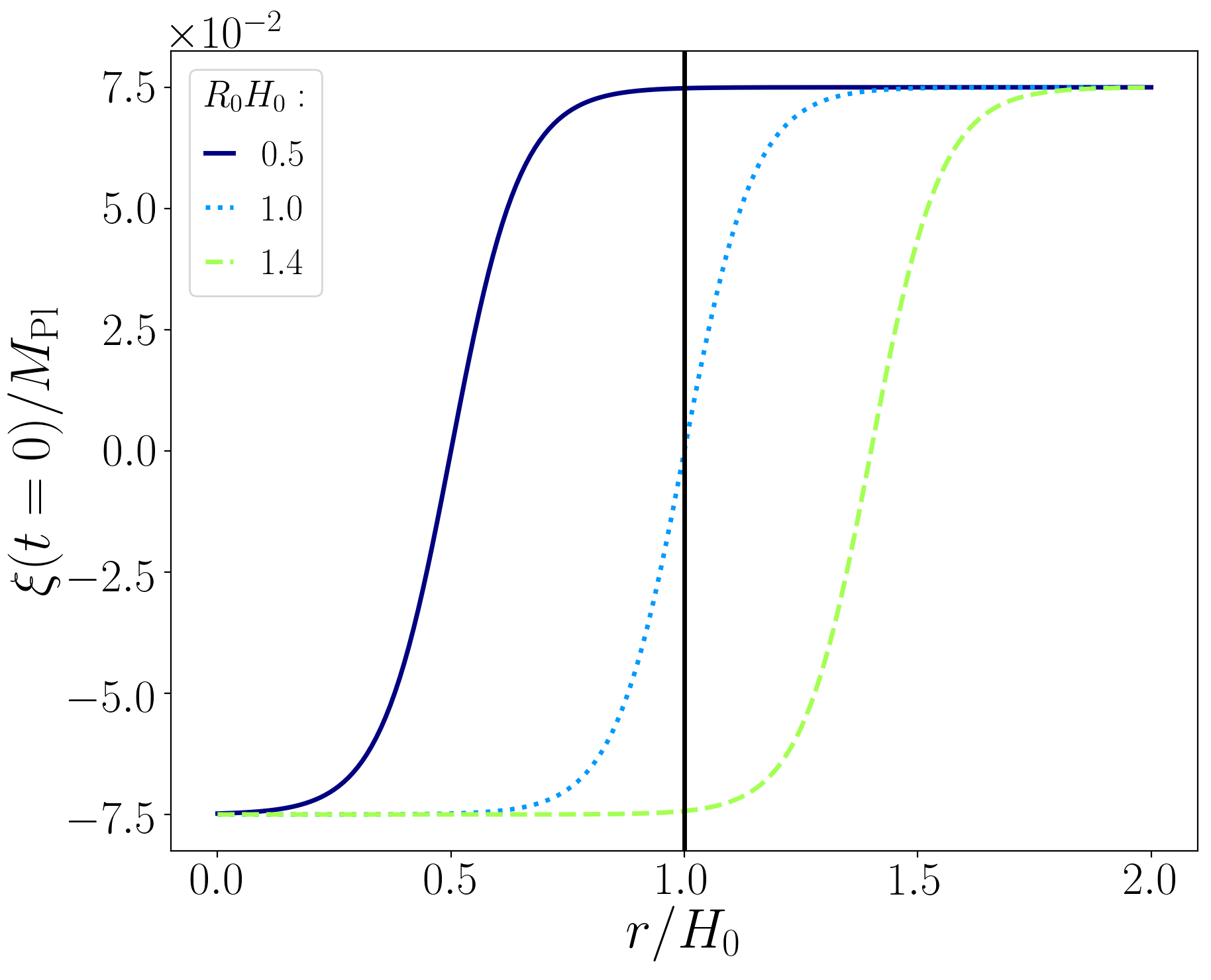}
    \caption{\textbf{Initial profiles} of the massless $\xi$ field, as a function of radial distance away from the center. Depicted are initially subhorizon, horizon and superhorizon sized perturbations, with same amplitude and thus same energy densities.} \label{fig::initial_profile_xi}
\end{figure}

This work was written based on simulations run using $\grchombo$ \cite{Clough:2015sqa}, with the CCZ4 formulation of the Einstein equations \cite{Alic_2012}. This formulation relies on the 4-dimensional spacetime being foliated into 3-dimensional non-intersecting hyperslices, whose intrinsic curvature is described by a spatial metric $\gamma_{ij}$, whilst their embedding in the 4-dimensional spacetime is encoded in the extrinsic curvature $K_{ij}$. The line element is decomposed as
\be 
    ds^{2} = -\alpha^2 dt^2 + \gamma_{ij}\big(dx^i + \beta^i dt\big)\big(dx^j + \beta^j dt\big) \,,
\ee
where the lapse $\alpha$ and the shift vector $\beta^i$ are user-specified gauge functions. The spatial metric $\gamma_{ij}$ is often written as a product of a  \textit{conformal factor} $\psi$ and a \textit{background} (or \textit{conformal}) metric $\bar{\gamma}_{ij}$, so that the determinant of the conformal metric equals one,
\be 
    \gamma_{ij} = \psi^4\bar{\gamma}_{ij}\,, \q \det{\bar{\gamma}_{ij}} = 1\,, \q \psi = \frac{1}{(\det{\gamma_{ij}})^{1/12}} \,.
\ee

Time evolution proceeds along the vector $t^a = \alpha n^a + \beta^a$, where $n^a$ is the unit normal vector to the hyperslice that is being evolved. The gauge functions $\alpha$ and $\beta^i$ are specified on the initial hyperslice, and then evolved using evolution equations suitable for long-term stable numerical simulations. The choice in this work is
\be
\begin{split}
    \partial_t\alpha &= -\mu_{\alpha_1}\alpha K + \beta^i \partial_i \alpha\,,\\
    \partial_t\beta^i &= \eta_{\beta_1} B^i\,,\\
    \partial_t B^i &= \beta^j\partial_j\bar{\Gamma}^i + \partial_t \bar{\Gamma}^i - \eta_{\beta_2}B^i\,,
\end{split}
\ee
where $\bar{\Gamma}^i\equiv \bar{\gamma}^{jk}\bar{\Gamma}^i_{jk}$, and $\bar{\Gamma}^i_{jk}$ are the conformal Christoffel symbols associated to the  conformal metric via the usual definition
\be 
    \bar{\Gamma}^{i}_{jk} = \frac{1}{2}\bar{\gamma}^{il}\big(\bar{\gamma}_{lk, j} + \bar{\gamma}_{lj, k} - \bar{\gamma}_{jk, l}\big)\,.
\ee
This choice is known as the \textit{moving-puncture gauge} \cite{Bona:1994dr,Baker:2005vv,Campanelli:2005dd,vanMeter:2006vi}. We choose $\mu_{\alpha_1} = 0.5,\, \eta_{\beta_1} = 0.75,\, \eta_{\beta_2} = 10^{-4}$, which helps us control constraint violation growth and evolve black hole spacetimes.

\subsection{Initial data} \label{section:initialdata}

\begin{figure}[t]
    \includegraphics[width=.915\linewidth]{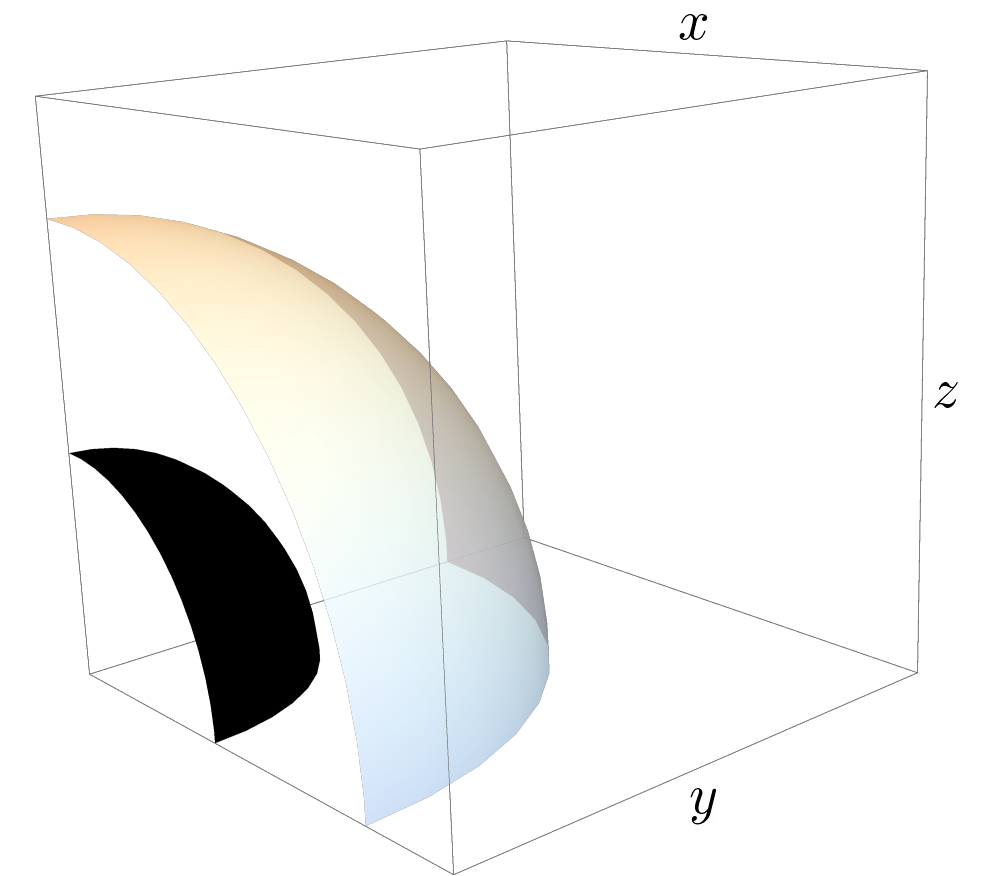}
    \caption{\textbf{Initial setup} of collapsing superhorizon overdensity (transparent shell) and inner initial Hubble horizon depicted in black. Using symmetric (reflective) boundary conditions we simulate an eighth of the system.} \label{fig::initial_setup}
\end{figure}

The matter content of this work is comprised by a massive $\phi$ field that dominates the background dynamics, and an inhomogeneous massless $\xi$ field which provides the local overdensity through its gradients. We choose potentials
\be 
    V_\phi(\phi) = \frac{1}{2}m^2\phi^2, \qquad V_\xi(\xi) = 0\,,
\ee
and spherically symmetric initial field configurations
\be 
\begin{split}
    \phi(t = 0, x^i) &= \phi_0
    \,,\\
    \xi(t = 0, x^i) &= \Delta\xi\tanh{\Big[\frac{r - R_0}{\sigma_0}\Big]}
    \,,\\
    \frac{\partial\phi(t = 0, x^i)}{\partial t} &= \frac{\partial\xi(t = 0, x^i)}{\partial t} = 0\,.
\end{split}
\ee
Examples of initial field profiles for the massless $\xi$ field are given in Fig. \ref{fig::initial_profile_xi}. The hyperbolic tangents allow us to localize the gradients in the $\xi$ field, and thereby its energy density, as the field is without potential. In our simulations, $mH_0^{-1} = 62.6$. We also choose a conformally flat metric $\bar{\gamma}_{ij}=\delta_{ij}$, so the energy density on the initial hyperslice is given by
\begin{equation}
    \rho(t = 0, x^i) = \frac{\psi^{-4}}{2}\delta^{ij}\partial_i\xi\partial_j\xi  
    +\frac{1}{2}m^2\phi_0^2\,,
\end{equation} 
which corresponds to a shell-like overdensity in a dark matter environment.

\begin{figure}[t!]
    \includegraphics[width=\linewidth]{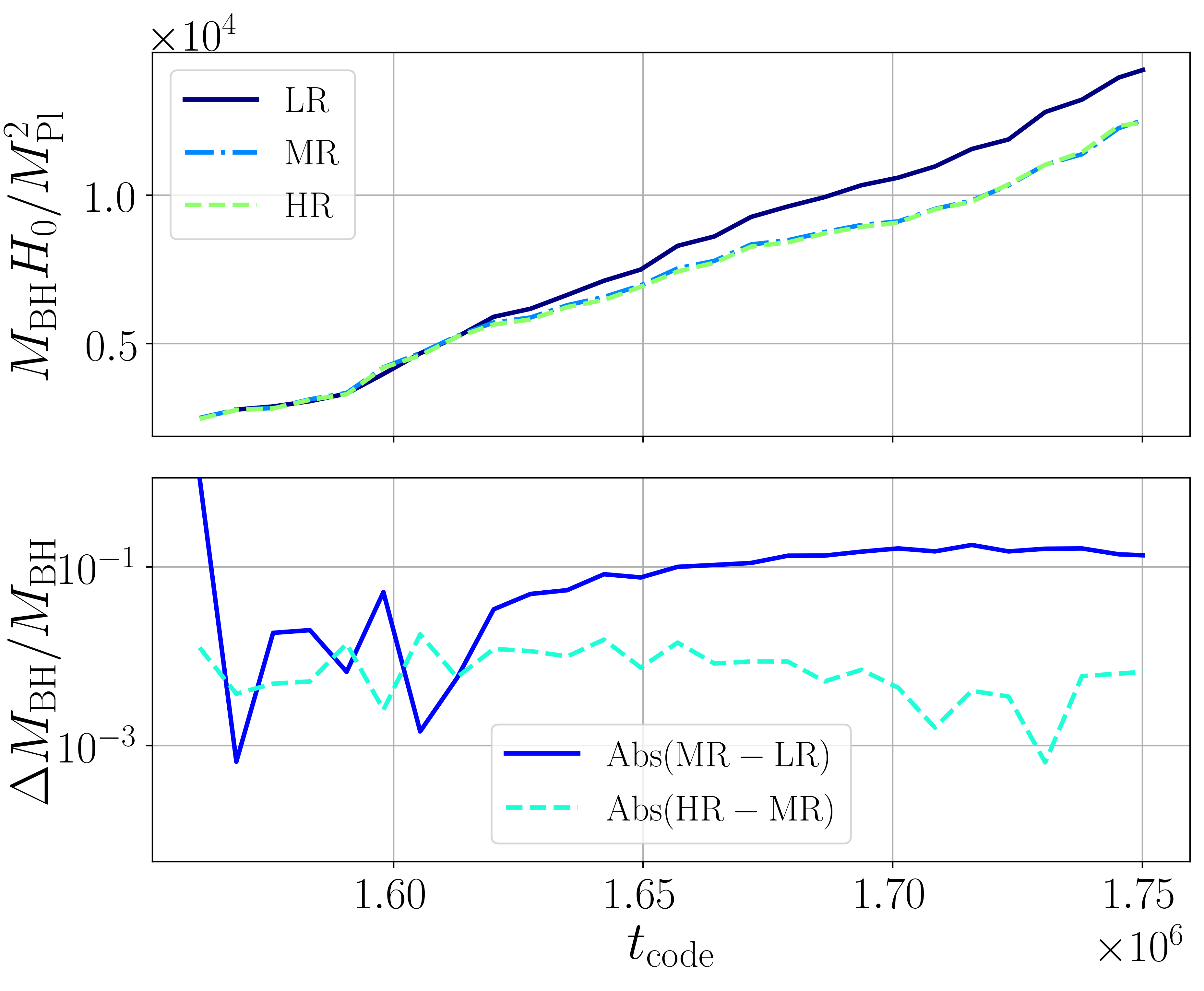}
    \caption{\textbf{Convergence test} of the black hole mass formed from an initial perturbation of $R_0=1.6H_0^{-1}$ and $\Delta\xi\mpl^{-1} = 0.075$. Top panel shows the estimated mass for three base grid resolutions $N_\mathrm{LR}=80$, $N_\mathrm{MR}=96$, $N_\mathrm{HR}=128$. Bottom panel shows errors in mass measurements between high-middle and middle-low resolutions showing convergence to $1\%$.} \label{fig::convergenceR1.6}
\end{figure}

This setup is spherically symmetric and we reduce the computational cost of evolution by simulating one eighth of the system using symmetric boundary conditions. A schematic depiction of the initial setup is given in Fig. \ref{fig::initial_setup}.

The equation of motion for a massive homogeneous field $\phi$ is given by the Klein-Gordon equation
\be \label{KGeom}
    \ddot{\phi} + 3H\dot{\phi} + \frac{\partial V(\phi)}{\partial\phi} = 0\,,
\ee
where $H\equiv \dot{a}/a$ is the Hubble function defined with the scale factor of the universe $a(t)$.

For the universe to stay matter dominated for an extended period of time, we require the solution of \eqn{KGeom} to be an undamped oscillation. This constrains the initial value of $\phi$ to
\be 
    \phi_0 < \frac{1}{\sqrt{3\pi}}\mpl\,,
\ee
and we use $\phi_0 = 7.8 \times 10^{-3}\mpl$ for all our simulations. Momentum constraints are trivially satisfied, and we solve the Hamiltonian constraint \eqref{eq:ham_constraint} numerically to find the conformal factor $\psi$.
\vspace{20pt}
\section{Convergence testing} \label{appendix::2}

We test the robustness of our numerical results by finding the mass of the black hole formed from an initial perturbation of radius $R_0=1.6 H_0^{-1}$ and $\Delta\xi= 0.15H_0^{-1}$, using three different base grid resolutions, namely $N_\mathrm{LR}=80$, $N_\mathrm{MR}=96$, $N_\mathrm{HR}=128$. Fig. \ref{fig::convergenceR1.6} shows the mass obtained with an apparent horizon finder \cite{Thornburg:2003sf} for these three runs, indicating that convergence is achieved.

Additionally, we made sure the code reproduces the FLRW limit for appropriate initial conditions. If one uses the same initial setup as described in appendix \ref{section:initialdata}, but gives the field $\xi$ a uniform value throughout the simulation box, evolution should proceed in an FLRW manner, as the simulated universe is now completely homogeneous. We checked this by averaging the values of the energy density $\rho$ and the scale factor $a$ over the box, and tracking these throughout the evolution. We satisfactorily find that the scaling between these two quantities is then $\rho \sim a^{-3}$, as expected for a matter-dominated FLRW universe.

\end{document}